\documentclass{elsart}
\usepackage[]{epsfig}
\usepackage{pstricks}
\usepackage{color,fancyhdr,graphicx,amsmath}
\begin{document}
 
\begin{frontmatter}

 \title{Maxwell's equal area law for charged Anti-deSitter black holes}
 \author{Euro Spallucci\thanksref{infn}}
\thanks[infn]{e-mail address: spallucci@ts.infn.it }
\address{Dipartimento di Fisica Teorica, Universit\`a di Trieste
and INFN, Sezione di Trieste, Italy}
 
\author{Anais Smailagic\thanksref{infn2}}
\thanks[infn2]{e-mail address: anais@ts.infn.it }
\address{Dipartimento di Fisica Teorica, Universit\`a di Trieste
and INFN, Sezione di Trieste, Italy}
       
 \begin{abstract}
 In this paper we present the construction of the  Maxwell
        equal area law in the \emph{Hawking temperature graph} for a charged
 black hole in Anti-deSitter background. We are able to find \emph{exact}
        solution for the corresponding \emph{isotherm} and entropies for
       "~gaseous~" (large) black holes and "~liquid~" (near-extremal)
         black holes.
        Isothermal construction removes
        the unphysical, negative heat capacity, regions. Furthermore,
         extremal black holes turn out to be dual to \emph{un-shrinkable} molecules of
         Van der Waals real fluid, which may explain their thermodynamical stability.
 \end{abstract}
 \end{frontmatter}
 \section{Introduction}
         Black hole (BH) thermodynamics is an example  of non-perturbative, semi-classical, quantum gravity effect
         based on the key results by Bekenstein and Hawking. This approach contained an implicit and
         not fully appreciated idea of Fluid/Gravity duality,
         \cite{Ambrosetti:2008mt,Bredberg:2010ky,Hubeny:2010wp,Hubeny:2011hd}
         which is one of the main outcome of the recent development in 
         $AdS/CFT $ duality \cite{Maldacena:1997re,Witten:1998zw,Witten:1998qj}. 
         The assignement of an entropy and a
         temperature to BHs assumes the existence of a (large) number of microscopic ``constituents'' (entropy)
         with an average kinetic energy (temperature).  The true nature of these constituents
         is still not well understood though a macroscopic description is available in terms of a generalized
         thermodynamics. In this description  the mass plays the role of internal energy, the area of the event
         horizon is identified with the entropy, while the Coulomb potential on the BH surface and the angular velocity
         are treated  as a sort of chemical potentials. \\
         One of the most important outcome of this approach is the discovery of phase transitions
         for BHs in AdS background \cite{Hawking:1982dh}. This result was extended to charged BHs
         in \cite{Chamblin:1999tk,Chamblin:1999hg}, where an analogy with the
         Van der Waals (VdW) description of a liquid-gas phase transition was described.
         While intriguing, this analogy relies mainly on a  shape similarity between the graph of the
         inverse Hawking temperature and the VdW isotherm in the P-V plane, thus forcing
         the identification of $1/T_H$ with $P$ and the BH radius $r_H$ with the volume. However, a proper
         definition of the canonical variables $\left(\, P\ ,V\,\right)$ was missing in the
         original formulation of the BH thermodynamics. \\
         Only recently, the cosmological constant has been promoted to the role of pressure and
         included in an extended version of the BH  First Law 
         \cite{Caldarelli:1999xj,Kastor:2009wy,Dolan:2010ha,Dolan:2011jm,Dolan:2012jh,Cvetic:2010jb,Lu:2012xu}.
         This extended approach allows to investigate AdS BH phase transitions in the P-V plane and give
          the VdW picture a non-ambiguous formulation \cite{Kubiznak:2012wp,Gunasekaran:2012dq}.\\ 
          A brief comment on the problems encountered  in the original formulation BH thermodynamics can help
          to understand the goal of this paper.\\  The whole picture of BH radiation is formulated
          as a semi-classical approximation where matter is quantized, but gravity is represented by a classical
          background geometry. On the other hand, the late stage of evaporation involves Planck
          scale physics implying that the semi-classical approximation becomes  unreliable
          in this regime. The problem can be by-passed for charged and rotating BHs where conserved charge, or
          angular momentum, stop the evaporation process when an extremal configuration is reached.  
          Even by assuming the most favorable situation, when charge and angular momentum are neither totally
          radiated away during the collapse ``balding-phase'', nor by quantum mechanical instability after the Bh forms,
           another problem remains standing. A charged AdS BH geometry is characterized by two
           length scales: the AdS curvature radius $l$, and the charge $Q$ (in geometric units).   
           Below a critical value of the ratio $Q/l$, the graph of the temperature as a function
           of the horizon radius develops an ``~oscillating~'' part corresponding  to a negative specific heat. 
           At the end-points of the oscillation interval (~which are a maximum and a minimum of the temperature~)
           the specific heat diverges.
           A negative specific heat implies the absence of thermodynamical equilibrium
           between the BH and the surrounding heat bath,  while divergences  signal the onset of a phase transition       
           where the physics of the system undergoes an abrupt change. \\
           In the absence of a proper understanding of the BH microscopic structure, no plausible physical picture of
           this behavior have been given so far. \\
           On the other hand, it is known that the VdW theory is a \emph{phenomenological} description of
            a "~real~" gas  which   breaks down below the critical temperature. In this regime, isotherms
            in the $P$-$V$ plane start to develop pair of extremal points.
            Between the minimum and the maximum an increase of volume corresponds to an unexpected increase of
            pressure which is in sharp contradiction with the experimental results indicating the presence of an
            \emph{isobaric plateau}.  The theoretical prediction and experimental data are reconciled
             by replacing the oscillating part of the isotherms by an
            horizontal isobar satisfying Maxwell's equal area construction.\\
            The Hawking temperature for an AdS charged BH shows a similar oscillatory behavior with the
            negative specific heat and thermally unstable, region. \\
             Then, one
            can implement the same Maxwell construction for the oscillating part of the BH temperature in (T,S) plane.\\
            In this paper we present the construction of the isotherm area-law in the Hawking temperature as a function
            of the entropy. This provides a 
            picture of the AdS BH critical behavior with a novel
            motivation  for the stability of  extremal BH.

  \section{Hawking temperature and the black hole VdW analogy}
 
         Our theoretical laboratory where to investigate the VdW fluid/BH duality is provided by the
         Reissner-Nordstr\"om Anti-deSitter (RNAdS) metric. The choice of the metric is suggested
         by its central role in $AdS/QCD$ duality in which the charged BH is  dual to a strongly coupled
         gluon plasma \cite{Maldacena:1997re,Witten:1998zw,Witten:1998qj,Myers:2008fv}.
         On the other hand, this model is simple enough for the reader to be able to follow the reasoning presented
         in the paper. The line element of the RNAdS metric is given as
          \begin{equation}
  ds^2=-\left(\, 1 -\frac{2M}{r} + \frac{Q^2}{r^2} +\frac{r^2}{l^2}\,\right) dt^2 +
          \left(\, 1 -\frac{2M}{r} + \frac{Q^2}{r^2} +\frac{r^2}{l^2}\,\right)^{-1} dr^2 + r^2\, d\Omega^2_{(2)}
 \label{uno}
 \end{equation}
         where, $M$ is the mass, $Q$ is the electric charge in geometric units, and the AdS curvature radius $l$
         is related to the cosmological constant
         as $\Lambda =-3/l^2$.  We briefly review the properties of the solution (\ref{uno})
         which are  relevant for our discussion. \\
          Metric (\ref{uno}) describes a non-degenerate black hole for a mass $M>M_0$ where $M_0$ is the  mass
          of an extremal configuration of radius $r_0$:
 
         \begin{eqnarray}
         && M_0=\frac{r_0}{3}\left(\, 2 + \sqrt{1+\frac{12Q^2}{l^2}}\,\right) \\
         && r_0^2=\frac{l^2}{6}\left(\, \sqrt{1+\frac{12Q^2}{l^2}} -1\,\right)
         \end{eqnarray}
 
\begin{figure}[h!]
                \begin{center}
                \includegraphics[width=10cm,angle=0]{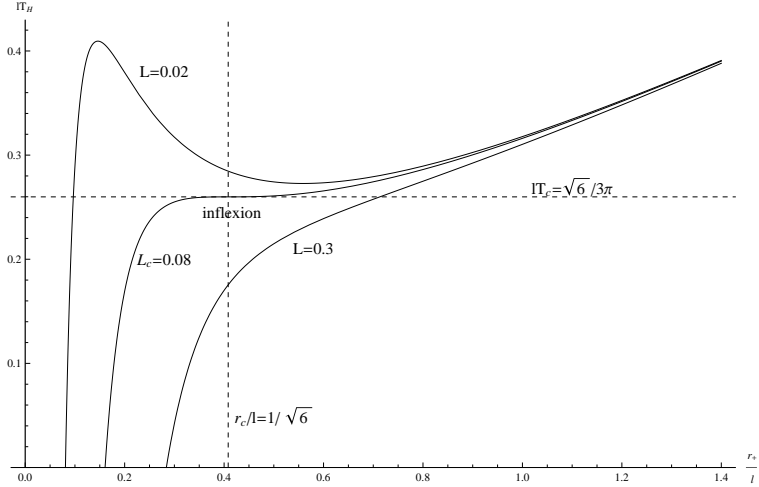}
                \caption{ Plot  of $T_H=T_H\left(\, r_H\,\right)$, for different values of $L^2=3Q^2/l^2$.
                At the critical value $L^2=1/12\approx 0.08$ the minimum and the maximum merge into an inflexion
                point at $r_c= l/\sqrt{6}$. $T_c=\sqrt{6}/3\pi l$ is the VdW critical temperature.
                   }\label{TH}
                  \end{center}
                 \end{figure}

           The corresponding Hawking temperature $T_H$ is given by (see Figure(\ref{TH})
          \begin{equation}
           T_H = \frac{1}{4\pi \, r_H}\left(\, 1-\frac{Q^2}{r_H^2} +\frac{3r^2_H}{l^2}\,\right)
          \label{ht}
           \end{equation}
       
          One can check that $T_H\left(\, r_0\,\right)=0$, as it is expected for any extremal configuration.\\
          The analysis of the temperature graph starts from the evaluation of the maximum and the minimum as
           
              \begin{equation}
             \frac{dT_H}{dr_H}=0 \longrightarrow
               \langle\, r_H^2\,\rangle _\pm= \frac{l^2}{6}\left[\, 1 \pm \sqrt{1 -\frac{36 Q^2}{l^2}}\,\right]
              \end{equation}
              The existence of  extrema  depends on the ratio between the charge $Q$ 
               (in geometric units)  and the AdS curvature radius $l$. \\
               For $Q^2 < l^2/36$,  there are a local \emph{maximum} at $r_- $ and  a local \emph{minimum} at $r_+$.
              The temperature graph for RNAdS has the same behavior as the temperature of the regular BH recently described
              in (\cite{Smailagic:2012cu}). Thus, it exhibits the same phase structure which can be summarized as
              follows:\\
              
              \begin{itemize}
               \item
               for $0\le T_H < T_{min} $ a small, thermally stable\footnote{By ``thermally stable'' we mean that
               the temperature decreases as $r_H$ decreases, and vice-versa.}, near-extremal black hole exists.
               \item
               for $T_{min}\le T_H < T_{max} $ there are:
             
              \begin{description}
\item[i] one small,(locally) stable, near-extremal black hole\\
              \item[ii] another unstable, intermediate black hole\\
              \item[iii] large, (locally) stable, black hole.\\
              \end{description}
              \item when $T_H>T_{max}$  there is one large, stable black hole.
              \end{itemize}
             
  The specific heat of the black hole is given by
                $C\equiv \frac{dM}{dr_H} \,\left(\,\frac{dT_H}{dr_H}\,\right)^{-1}$
                 and diverges at the extrema. It is \emph{negative}
                for $r_{max}< r_H < r_{min} $. The negative sign means that the BH
                cannot  establish a thermal equilibrium with the environment.\\
                Our poor understanding of the BH microscopic structure
                does not permit to grasp true physical nature of these changes of state.\\
                
               These extrema merge into an \emph{inflexion} point for $ Q^2=l^2/36 $, corresponding to
               the critical values given by
              \begin{equation}
               Q^2=\frac{l^2}{36}\longrightarrow  \langle\, r_c^2\,\rangle = \frac{l^2}{6}=\sqrt{6}\, Q
              \end{equation}
              and
              \begin{equation}
               T_c\equiv T_H\left(\, r_c\,\right)=\frac{\sqrt{6}}{3\pi\, l}=\frac{\sqrt{6}}{18\pi\, Q}
              \end{equation}
             
               On the other hand, when $Q^2 > l^2/36$, $T_H$ is \emph{monotonically increasing}
               and there is a single, thermally stable black hole. \\
               The temperature graph analysis includes only the temperature and the geometric volume  $V=4\pi\,r^3_+$  as
              thermodynamical variables,while the cosmological constant $\Lambda$ is simply a free parameter such as $Q$.\\
               An early attempt to introduced the \textbf{pressure} was carried out in \cite{Chamblin:1999tk,Chamblin:1999hg}.
               It was pointed that, plotting $1/T_H$ versus $r_+$, the resulting graph looks alike
               the VdW isotherm \textbf{provided} one identifies $1/T_H$ with $P$, $r_+$ with the volume $V$,
               and assigns $Q$ the role of ``temperature''. While interesting, this approach
               is not uniquely defined and mismatches intensive and extensive thermodynamical
               variables \cite{Kubiznak:2012wp}. \\
               Recently, an alternative VdW description of BH phase transitions has been presented
               in \cite{Kubiznak:2012wp,Gunasekaran:2012dq} where the cosmological constant is
               assigned the role of the pressure for a ``real gas''
               \cite{Caldarelli:1999xj,Kastor:2009wy,Dolan:2010ha,Dolan:2011jm,Dolan:2012jh,Cvetic:2010jb,Lu:2012xu},
               while the specific volume is identified with $v=2l_P^2\,r_H$,
               where $l_P$ is the Planck length. The temperature of the
               gas is  the Hawking temperature.
               This interpretation is much more physically sound and avoids the above mentioned confusion among intensive and
               extensive variables. The key point is to introduce in the equation (\ref{ht})  two thermodynamical variables
               \begin{eqnarray}
               && P= -\frac{\Lambda}{8\pi L_P^2}\ ,\quad\hbox{pressure}  \label{press}\\
               && \mathcal{V}= 2r_H L_P^2\ ,\quad\hbox{specific  volume}  \label{vspec}
               \end{eqnarray}
               While the cosmological
               constant is the most likely candidate for the pressure,\footnote{   As a matter of fact, 
               one simply has to switch  from the  usual
               interpretation of $\Lambda$ as the  "~vacuum pressure~" , to a classical picture, where $\Lambda$ is the pressure
               of a "~real gas~".}, it is less intuitive  to identify the specific volume with the radius of the horizon.\\
               Once the equation (\ref{ht}) is written in the form
                $P=P\left(\, \mathcal{V}\,\right)$, its graph
                 looks like the VdW  curve for a ``real gas'' at  the temperature $T_H$:
                
                 \begin{equation}
                  P=\frac{T_H}{\mathcal{V}}-\frac{1}{2\pi \mathcal{V}^2} +\frac{2Q^2}{\pi \mathcal{V}^4}
                 \label{VdW}
                 \end{equation}

                Now, the same equation can be seen either as the Hawking temperature for a BH of radius
                $r_+$ in a given, charged, AdS background (\ref{ht}) , or as the pressure of a Van der Waals fluid (\ref{VdW})
                enclosed in a (specific)volume $\mathcal{V}$ and heated at the temperature $T$.\\
                \emph{The two systems  are dual}\footnote{  The term ``~duality~" is often used with very different meanings in
                 the literature. The duality we introduced in this paper is very similar to the string theory T-duality, where
                 the \emph{same} equation is interpreted either as the spectrum of a closed string wrapped around a compact dimension
                 of radius $R$, or the spectrum of
                 a closed string wrapped around a compact dimension of radius $\alpha^\prime/R$.
                 In our case exchanging the role of $P$ and $T$
                 is the analogue of exchanging Kaluza-Klein and winding modes in string theory. If one is used to see duality 
                as an inversion of a coupling constant, in our case the corresponding constant is Boltzman constant
                $k\rightarrow 1/k$. }
                \emph{to each other in the sense that they are described
                by  the \emph{same} equation. The mapping between the two systems is given by
                equations (\ref{press}),(\ref{vspec})}.\\
                For further discussion it is useful to recall that the VdW  theory is a
                "~phenomenological~" description of a real gases.
                 Below the critical temperature the VdW formula predicts an "~oscillatory~" region characterized
                 by a pair of extremal points. This
                 behavior is in contradiction with the experimental evidence of an "~isobaric plateau~",
                 where the pressure remains constant while the specific volume changes.\\
                 In order to reconcile  theory and experiment Maxwell introduced the ``~equal area law~''.
                 This is a prescription aimed to replace the oscillatory region with an isobar,
                 $P=P^\ast$,  cutting the pressure graph in such a way that the area below and above
                 the isobar are the same:
                 \begin{equation}
                  \int_l^g  \mathcal{V}dP =0\longrightarrow P^\ast\left(\, \mathcal{V}_g -\mathcal{V}_l\,\right)=
                  \int_{\mathcal{V}_l}^{\mathcal{V}_g} Pd\mathcal{V}
                 \label{plaw}
                  \end{equation}
                  where $P=P^\ast$ denotes the \emph{equal area isobar}.\\
                  The physical interpretation of the constant pressure part of the curve is that it
                  describes the  coexisting, mixed phase of fluid and gas.\\
                  Equations (\ref{VdW}) and (\ref{plaw}) define $P^\ast$, $\mathcal{V}_l$, $\mathcal{V}_g$. So far, only
                  perturbative solutions, around critical values, are known. In the next section we shall give  a new,
                  \emph{non-perturbative}, derivation of  $\mathcal{V}_l$, $\mathcal{V}_g$ and discuss a new picture
                  of the BH phase transition.

                  \section{The Maxwell construction for the charged AdS black hole in (T,S) plane.}
 
                   In the second part of this paper, we will show that the kind of duality defined above allows to
                  find \emph{exact} solutions in  the $\left(\, P\ ,\mathcal{V}\,\right)$ plane.\\
                  Paralleling the gas behavior to the BH thermal instability along the oscillatory part of the Hawking
                  temperature graph,
                   invokes the same "~\emph{flattening}~"
                  procedure in the equation (\ref{ht}). \\ The key idea of this paper is to complete the physical analogy
                   between the AdSRN BH and the VdW fluid by
                   implementing the \emph{Maxwell construction in the gravitational sector} and  exchanging
                   the isobar equal-area, in the gas picture, with an isotherm in the BH picture.\\
                   Equal-area law is obtained starting  from the  the BH free-energy given by
 
\begin{equation}
F= M-TS  \label{onfree}
\end{equation}
 
Variation of (\ref{onfree}), at constant P, leads to
 
\begin{equation}
dF= -SdT  \label{dfree}
\end{equation}
 
Integrating the equation (\ref{dfree}), and keeping in mind that co-existing phases have the same free energy by definition,
one finds the \emph{equal area law} in the $(T\,S)$ plane as:
 
\begin{equation}
T^\ast\left(\, S_g -S_l\,\right)
=\int_{S_l}^{S_g} T\left(\, S\,\right)\, dS
\label{t1}                
\end{equation}

The Hawking temperature as a function of the entropy and pressure reads
 
\begin{equation}
T=\frac{1}{4\sqrt{\pi S}}\left(\, 1 + 8 P S - \frac{\pi Q^2}{S}\,\right)
\label{thp}
\end{equation}
 
with the BH entropy given by $S=\pi r_H^2 $. It is convenient to introduce reduced variables as
 
\begin{eqnarray}
&& S= s S_c \ ,\qquad S_c\equiv 6\pi Q^2\ ,\\
&& T= t T_c \ ,\\
&& P= p P_c \ ,\qquad  P_c\equiv \frac{1}{96\pi Q^2}\\
&& \mathcal{V}=v  \mathcal{V}_c                
\end{eqnarray}
 
Thus, the Hawking temperature can be written in the form
 
                 \begin{equation}
                    t=\frac{3}{4\sqrt{s}}\left(\, 1 + \frac{1}{2} p s - \frac{1}{6s}\,\right)
                   \label{tredux}
       \end{equation}
and is valid for any value of the charge $Q$.
The equation(\ref{t1}) gives the Maxwell isotherm as
 
                \begin{equation}
  t^\ast= \frac{1}{\sqrt{s_g}+\sqrt{s_l}}\left(\, \frac{3}{2}-\frac{1}{4\sqrt{s_l s_g}} +
           \frac{p}{4}\left(\, s_l + s_g + \sqrt{s_l s_g} \,\right)  \,\right)
                    \label{t2}                  
      \end{equation}
 
where $s_l,\,s_g$ correspond to the entropy of liquid and gaseous phase, respectively.

               \begin{figure}[h!]
                \begin{center}
                \includegraphics[width=10cm,angle=0]{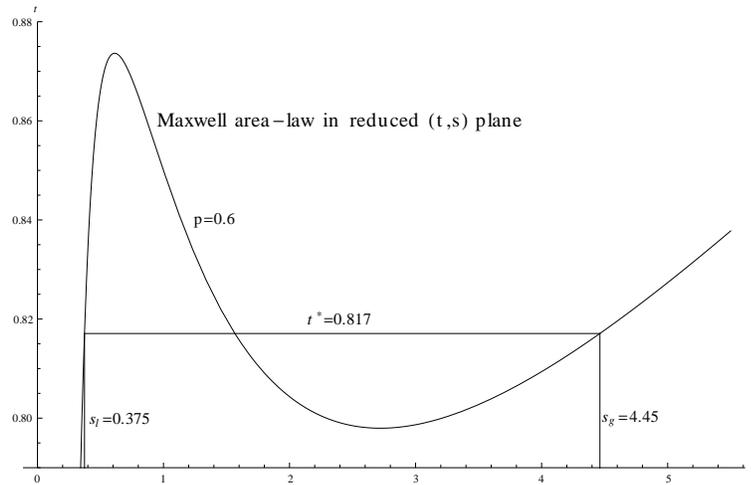}
                \caption{ Graph of rescaled temperature below critical reduced pressure\,$p< p_c=1$.
                 The specific heat is negative between two extrema. The equal area isotherm construction
                   is shown. }\label{BHT4}
                  \end{center}
                 \end{figure}

Equations (\ref{t2}), (\ref{tredux}) determine the solutions for $s_l$, $s_g$.
Keeping in mind that $s_l\,\,\,s_g$ correspond to the same \emph{isotherm}, one finds from  equation (\ref{tredux})
 
\begin{eqnarray}
 t(s_l)&=&t(s_g)\nonumber\\
0&=& 1-\frac{1}{2}p\,x-\frac{1}{6}\frac{y^2-x}{x^2}\nonumber\\
 x&=&\sqrt{s_l\,s_g}\nonumber\\
 y&=& \sqrt{s_l}+\sqrt{s_g}\nonumber\\
 \sqrt{s}_{l,g}&=&\frac{1}{2}\left(\, y \pm \sqrt{ y^2 -4x}\,\right)\nonumber\\
y&=&-3p\,x^3+6x^2+x
\label{y}
\end{eqnarray}
On the other hand, the area law (\ref{t2})leads to
\begin{eqnarray}
t^\ast&\equiv& \frac{t(s_1)+t(s_2)}{2}=\frac{y}{16\,x^3}\left(3p\,x^3+6x^2+3x-y^2\right)\nonumber\\
t^\ast&=&\frac{1}{4x\,y}\left(-p\,x^2+6x-1+p\,x\,y^2\right)
\label{x}
\end{eqnarray}
Combining equations (\ref{y}) and (\ref{x}), we get
 
\begin{equation}
  p^2\,x^4-2p\,x^3+2x-1=0
\end{equation}
 
This equation can be solved analytically by rewriting it in the form
 
\begin{eqnarray}
  &&0= x^2\left(p\,x-1\right)^2-\left(x-1\right)^2\nonumber\\
 && x_{1,2} =\pm \frac{1}{\sqrt{p}}\qquad\longrightarrow y^2=\frac{2}{p}\left(\, 3 -\sqrt{p}\,\right)\\
 && x_{3,4} =\frac{1}{p} \left(\, 1\pm\sqrt{1-p}\, \right)\longrightarrow y^2= 4x
\end{eqnarray}
 
Physically, only $ x_{1,2}$ are acceptable because $x_{3,4}$ lead to a reduced entropy larger than one.
 Finally, solutions for the entropies $s_{l,g}$  and the isotherm $t=t^\ast$ in the $\left(\, p\ ,s\,\right)$  plane are given by

\begin{eqnarray}
  s_{l,g}&=&\frac{1}{2p}\left(\sqrt{3-\sqrt{p}}\pm\sqrt{3-3\sqrt{p}}\right)^2\\
  t^\ast&=&\sqrt{p\left(3-\sqrt{p}\right)/2}\label{st}
\end{eqnarray}
 
\begin{figure}[ht!]
                \begin{center}
                \includegraphics[width=10cm,angle=0]{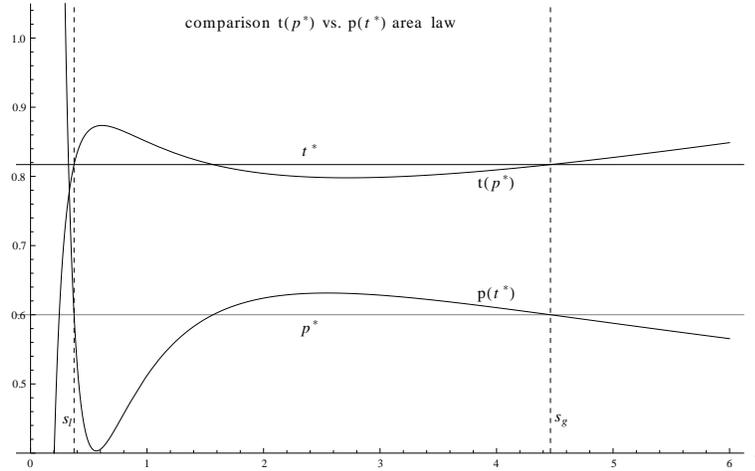}
                \caption{In his figure we plot together $p=p(s)$ and $t=t(s)$ in order to show that the equal areas in
                 both cases correspond to the same $s_{l,g}$. }
                \label{twoplots}
                  \end{center}
                 \end{figure}

Thus, the oscillating part in Figure( \ref{BHT4}), below the critical pressure $p_c$,
should be replaced with the isotherm $t=t^\ast$ in the same way as it is done in the
 VdW case.\\
\\
Correspondence between $\left(\, p\ ,v\,\right)$  and $\left(\,t\ ,s\,\right)$ plane, in view of our result, 
goes far beyond a simple matter of taste in which plane one wants to work. The correspondence between
the two descriptions is depicted in a following scheme and plotted in Fig.(\ref{twoplots})
 
\begin{center}
  \begin{tabular}{|c|c|c|}\hline
  $\left(\,t\ ,s\,\right)$ plane &              & $\left(\, p\ ,v\,\right)$ plane\\ \hline
  $t\left(\,s\,\right)_{p=p^\ast}$\, t-graph & $\rightarrow$ & $p\left(\, v\right)_{t=t^\ast}$\, p-graph\\ \hline
  $t=t^\ast(p^\ast)$\,\, equal-area isotherm & $\rightarrow$ & $p=p^\ast\left(\,t^\ast\,\right) $\,\,equal-area isobar\\
\hline
  \end{tabular}
  \end{center}
 
Now, we are ready to introduce explicit solutions in the $\left(\, p\ ,v\,\right)$ plane as well.
Inverting equation (\ref{st}) gives the equation
 
\begin{equation}
 x^3 -3x^2 +2t^2 =0
\end{equation}
 
where $x=\sqrt{p}$, which has solutions
 
\begin{eqnarray}
&& x_1 = 1-2\cos(\frac{\phi}{3})\ , \\
&& x_2 =1-2\cos(\frac{\phi-\pi}{3}) \ ,\\
&& x_3 =\sqrt{p^\ast(t)}=1-2\cos(\frac{\phi+\pi}{3})\ ,\\
&& \cos\phi= 1-t^2 \nonumber
\end{eqnarray}
 
$x_3$ is the physically interesting solution since it gives $p\le 1$. Therefore, exact solutions in reduced
$\left(\, p\ ,v\,\right)$ plane are
\begin{eqnarray}
  v_{l,g} &&=\frac{\sqrt{1+\cos\left(\frac{\arccos(1-t^{\ast\,2})+\pi}{3}\right)}
  \mp\sqrt{3\cos\left(\frac{\arccos(1-t^{\ast\,2})
  +\pi}{3}\right)}}{1-2\cos\left(\frac{\arccos(1-t^{\ast\,2})+\pi}{3}\right)}\\
  p^\ast&&=\left[1-2\cos\left(\frac{\arccos(1-t^{\ast\,2})+\pi}{3}\right)\right]^2
\end{eqnarray}

\begin{figure}[ht!]
                \begin{center}
                \includegraphics[width=10cm,angle=0]{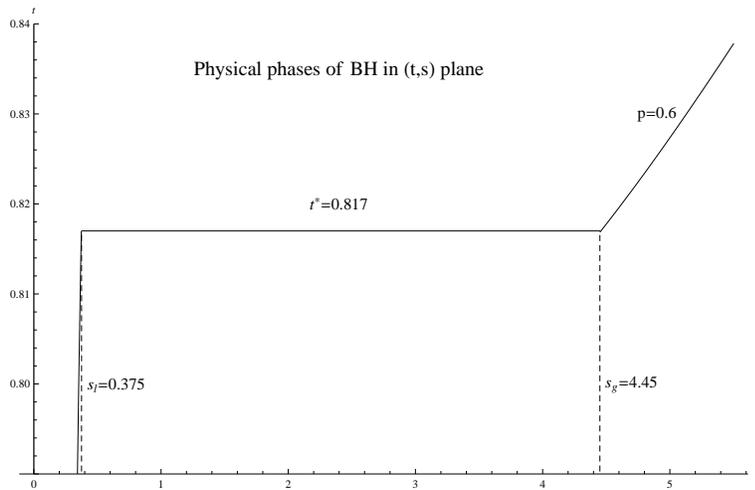}
                \caption{ Hawking temperature for ``real''BH above the critical pressure. The plateau
                  corresponds to the mixed phase while BH is changing from
                 the ``~gaseous~'' to the ``~liquid~'' phase.}\label{true}
                  \end{center}
                 \end{figure}

Maxwell construction eliminates the oscillating part of $T_H$ corresponding
to the negative specific heat region, which is analog of the unphysical part of the VdW curve.
The temperature graph of the physical BH phase transition is given by the Figure(\ref{true}).\\
 
             Thus, in our picture  the temperature graph of the physical BH phase transition should be
              as in Figure(\ref{true})
 
 The Maxwell construction modifies significantly the phase structure as described in the first part of the paper.
In the gravitational sector we have following  phases:
\begin{itemize}
\item for  $s_0 \le s\le s_l$ we find a single BH in a``~liquid-phase~'';
\item for $s> s_g$  we find a single BH in a ``~gas-phase~'';
\item for $s_l \le s\le s_g$ we find BHs  in a ``~coexisting gas-liquid-phases~'' transiting from one phase to another
\end{itemize}
 
The isotherm part of the graph, replacing the oscillatory region, represents the phase transition during which
the black hole is in the mixed state and undergoes gradual transition from the gaseous to liquid state.
This transition is characterized by a Hawking emission during which the radius of the
horizon is decreasing while temperature remains constant, contrary to the usual situation.
This presumably indicates some kind of microscopical constituents "condensation`` which remains to be better understood.
Once the radius $r_l$ is reached the black hole is in a ``~liquid-phase~''. \\
 
 One could be tempted to say that during the change of state, the cosmological constant (pressure) of the background geometry
is varying.  This, however, is not the case. The reason is that the Maxwell construction for the real gas with varying P
replaces the oscillating part of the \emph{constant} temperature curve with a \textbf{constant} pressure
segment.
It is only in this side of the duality that the question :``does the pressure vary during the phase transition?''
can be answered.
The analog of this isobaric plateau, in the gravitational case, is a \emph{constant} temperature interval at 
\emph{constant} pressure. During the
coexisting phases regime the BH evolution significantly departs from the Hawking behavior: the emission of thermal radiation
does not diminish the temperature and does not affect the cosmological constant. It is the BH ``internal structure''
which is suffering some sort of re-arrangement, not the background AdS geometry.\\
Finally, the BH reaches its  extremal configuration which is dual to the minimal  ``~molecular~'' volume for the VdW
fluid. This analogy provides an intuitive explanation for  the stability, i.e. in-compressibility, of the extremal BH.
It may seem that this type of stable ground state is limited to charged BHs,
but this is not the case as we discuss in the conclusions.
 
\section{Conclusions }
 
In this paper we have extended the idea of fluid/gravity analog in order to provide a new picture
of the isothermal behavior of critical charged BH in AdS background.
 We construct the Maxwell area law  in the Hawking temperature description of BH.
The outcome of this procedure is that  physical BH
undergoes an \emph{isothermal}  transition from gas to liquid phase at constant pressure. Consequently
there are neither regions with negative nor divergent specific heat.
Furthermore,  we were able to find analytic solutions the area law both
in $\left(\, T\ , S\,\right)$ and $\left(\, P\ , v\,\right)$ planes.
We conclude that working in the $\left(\, T\ , S\,\right)$ plane gives non-trivial advantage with respect to the
WdV description in (P,V) plane.\\
Last but not the least, duality identifies the minimal molecular volume with extremal BH configuration providing
a physical argument for its incompressibility. \\
It may seem that the described mechanism is limited to RNAdS, because of the
key role played by the electric charge in providing extremal configuration. We claim that the same behavior will occur in any
geometry endowed with a short-distance length scale, independently of its origin. 
We really have in mind a larger picture involving regular BHs that have been investigated in a recent set of papers
\cite{Nicolini06,Ansoldi:2006vg,Spallucci:2008ez,Nicolini:2008aj,Smailagic:2010nv,Nicolini:2009gw,Mann:2011mm,Mureika:2011hg}.
In these geometries a minimal length is embedded into the space-time fabric at the fundamental
level. What is crucial for duality to work
is the existence of an extremal configuration, notoriously known to exist for charged and/or rotating BHs in
standard geometries, but also present in the case of regular neutral, non-rotating BHs.
In a picture where extremal BHs are viewed as un-shrinkable molecules of gas, it is natural to look for
a relation between the size of these molecules and the minimal length from the vantage point
of the self-complete  quantum gravity approach  recently proposed in 
\cite{Dvali:2010bf,Dvali:2010ue,Spallucci:2011rn,Spallucci:2012xi,Nicolini:2012fy,Mureika:2012fq}.
The phase portrait for such BHs will be presented in a forthcoming paper. 
This note is limited to the RNAdS black hole for we believe
that readers are more familiar with this geometry and can grasp better our approach idea and its
consequences.

\end{document}